\documentclass[11pt]{article}
\usepackage{amsmath}
\usepackage{amsfonts}
\usepackage{graphicx}
\usepackage{hyperref}
\usepackage{url}
\usepackage{xcolor}
\usepackage{microtype}
\usepackage{booktabs}
\usepackage{fancyhdr}
\usepackage{lipsum}
\usepackage{siunitx}
\usepackage[T1]{fontenc}
\usepackage[a4paper, left=2.1cm, right=2.1cm]{geometry}
\title{Metros reduce car use in European cities but trams do not}

\author{Rafael Prieto-Curiel\thanks{Complexity Science Hub, Metternichgasse 8, Vienna, Austria. \texttt{prieto-curiel@csh.ac.at}} }

\date{}

\begin{document}
\maketitle

\section*{Abstract}

Despite the evident drawbacks, car ownership and usage continue to rise globally, leading to increased pollution and urban sprawl. As alternatives, Active Mobility and Public Transport are promoted for their health, economic, and environmental benefits. However, the efficiency of Public Transport varies widely. Metro systems, in particular, offer a high-capacity, long-distance solution, but they are expensive and only found in a limited number of cities. Trams, on the other hand, may serve as a substitute. This study compares the modal share in European cities, analysing the differences between those that have a metro, a tram, or neither. The analysis draws on a comprehensive dataset from \href{https://citiesmoving.com/}{CitiesMoving.com}, which compiles and harmonises mobility surveys from around the world according to the $ABC$ framework ($A$ for Active mobility, $B$ for Bus and other forms of Public Transport, and $C$ for Cars). Findings reveal that cities with a metro have a significantly lower share of car journeys than those with only a tram or no rail system.

\section{Introduction}

%%%% cars in cities are terrible %%%% yet, the number of cars in the world is growing yy
{
Cars impose a significant burden on society, broadly categorised into environmental issues, space consumption, lifestyle effects, and safety concerns \cite{miner2024car}. While much attention is given to fuel consumption, cars have a far more significant environmental impact due to their production, toxic materials (like tyres, brakes, asphalt), and waste disposal \cite{berrill2024comparing}. They also demand vast amounts of street space and costly infrastructure maintenance \cite{haberl2021stocks}, which has devastating impacts in cities \cite{aiello2024urban}. Moreover, cars are becoming bigger and heavier, requiring more materials for their manufacture \cite{cuenot2017wider}. Additionally, cars promote sedentary lifestyles, reducing the benefits of active mobility, such as social integration and street commerce \cite{yoshimura2022street}. Alarmingly, cars have caused at least 60 million deaths and 2 billion injuries, many of whom were walking or cycling during the crash \cite{miner2024car}. Yet, car ownership and use continue to rise. In Mexico, for example, car numbers surged from 10 million in 2000 to nearly 40 million by 2022 \cite{PrietoInventarioMovilidad}. Globally, while the population is projected to grow 24\% between 2015 and 2040, the number of cars is expected to double \cite{DoubleCars}. 
}

%%%% an alternative is active mobility and public Transport yy
{
Commuting is regarded as one of the least enjoyable daily activities, so people seek faster, more reliable, secure, or cheaper alternatives \cite{grubler2003technology, krueger2009national, prieto2022ubiquitous, van2010perceptions}. Therefore, to reduce car dependence, it is essential to explore viable options for commuting. Active mobility and Public Transport are more sustainable than cars. Walking and cycling for short trips offer mental and physical health benefits, boost local economies, and promote greener urban spaces \cite{pisoni2022active, nieuwenhuijsen2020urban, croeser2022finding}. Public Transport is a cost-effective, energy-efficient, and less polluting alternative that enhances social equity \cite{liao2020disparities, banister2011cities}. However, Public Transport is often perceived negatively and associated with unreliable or uncomfortable journeys \cite{de2016travel}. Additionally, many cities still prioritise car infrastructure over Public Transport and Active Mobility, often influenced by the car lobby \cite{MultimodalSzell2022, douglas2011cars}. 
}

%%% among the many forms of public Transport, metro is different
{
Most cities have some form of Public Transport, ranging from infrequent and informal minibuses to integrated networks of metro, trams, cable cars, and BRTs. Public Transport typically follows a hierarchical structure: buses and minibuses serve last-mile trips, while metros and BRTs handle long-distance travel \cite{wang2024quantification}. A key challenge for cities is choosing the most adequate transport mode \cite{yang2017choosing}. Should they invest in a metro, or are there more effective alternatives? A metro can move vast numbers of people efficiently and encourage nodal development around stations \cite{yang2017choosing}. They also reduce air pollution and greenhouse gas emissions \cite{lin2022metro}. However, metros are expensive to build and maintain, requiring high ridership to be viable. Globally, metros remain relatively rare, though their presence has grown, particularly in the Asia-Pacific region \cite{UITP2021}. A common alternative to a metro, especially in smaller urban areas, is light rail transit (or trams). In general, trams are cheaper to build but have lower capacity and speed than a metro. In many cities, trams serve as supplementary or last-mile connections to metro systems \cite{UITP2023}. And while metros and trams serve distinct roles, differing in speed, capacity, station density, and integration with urban infrastructure \cite{yang2017choosing}, they are sometimes viewed as substitutes. But are they genuinely interchangeable?

At their peak in 1930, around 900 cities operated trams (Figure \ref{Figure0}). However, competition with cars led to a steep decline in the mid-20th century, resulting in the shutdown of nearly two-thirds of the world's tram networks \cite{UITP2023}. Only recently have new tram systems emerged. 

\begin{figure}[h!] \centering
\includegraphics[width=0.65\textwidth]{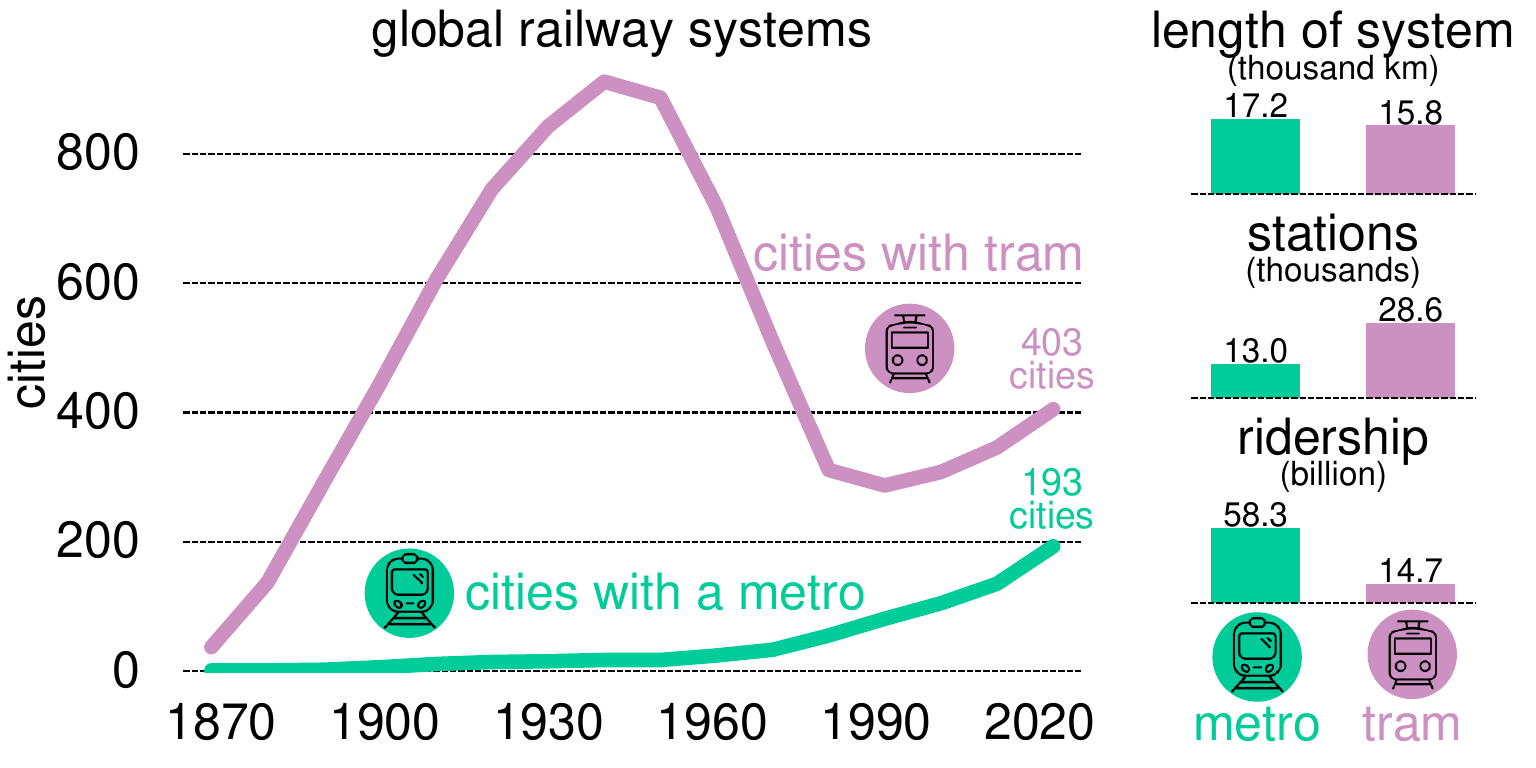}
\caption{Global railway systems by metro and tram coverage, length, stations, and ridership. Left - Number of cities with a metro and a tram system between 1870 and 2020. Right - Global length of the system, number of stations (or stops) and yearly passengers for metro and tram. Data from \cite{UITP2021} and \cite{UITP2023}.}
\label{Figure0}
\end{figure}
}

%%%% are they substitutes?
{
High-quality, efficient Public Transport systems attract users and help reduce car dependence \cite{redman2013quality, beirao2007understanding}. For example, it has been modelled that increasing service frequency is possible to increase passenger volume \cite{reinhold2008more}. And other elements, such as Public Transport speed and waiting time, have also been considered \cite{levinson2007planning}. However, at the city level, it remains unclear which specific characteristics of Public Transport are most effective in attracting users and in reducing car use, and how these dynamics vary with city size \cite{saeidizand2025car}. For instance, what level and type of Public Transport provisions are needed in a city of 100,000 residents compared to a city of one million? Are the requirements similar, or do they differ significantly? Thus, a key question is how commuting patterns differ in cities with a metro, a city that has only a tram but no metro, or neither. How does modal share change? And crucially, can a tram be as effective as a metro in reducing car use? When Public Transport is inadequate, it enters a vicious cycle, where the poor service leads to declining ridership, further reducing service frequency \cite{mogridge1997self, bar2013model}. As a result, many users switch to other modes, often cars \cite{prieto2021paradox}. But would a tram fall short as a viable commuting alternative?
}

%%%% here we
{
Here, the modal share across cities is analysed within the framework of the $ABC$ of mobility \cite{MobilityABCPrietoOspina}. The model categorises all Active Mobility modes of transport into $A$, all Public Transport into $B$, and all Cars into $C$, capturing their relative frequencies. Cities are then grouped according to whether they have a metro, only a tram, or neither, and the share of $C$ journeys is compared among these groups. Cities with a metro have a smaller car share than cities with only a tram or cities without a rail system. Cities of medium and large size with a metro have roughly half the Car share than cities without a metro, but this is not observed for cities with only a tram.
}

\section{Results}

%%%% data for the ABC for 380 cities in Europe, 
{
Data corresponding to the modal share of cities was obtained by combining mobility surveys conducted at the city level \cite{MobilityABCPrietoOspina}. The mobility data concentrates on metropolitan areas, gathered from many sources, including the European Platform on Mobility Management (EPOMM) and ICLEI, with data supplied by the EcoMobility Alliance Cities \cite{Wiersma2021, Duleba2021, GovernmentOfficeforScience2018}. The modal split has been categorised into three separate groups: $A$ for all types of Active Mobility (including walking and cycling), $B$ for all kinds of Public Transport (including metro, tram, buses and others) and $C$ for Car mobility (including cars, taxis, motorbikes, and mobility apps). The data represent daily commutes obtained from the latest mobility surveys of each city, most of which were conducted before the COVID-19 pandemic \cite{MobilityABCPrietoOspina}. There are some issues related to the mobility data that make comparisons across cities complex \cite{saeidizand2022revisiting}. For example, the parts of the city covered by the surveys, the purpose of journeys, city size and its demographic composition \cite{vanoutrive2015modal, yang2017choosing}. Additionally, some surveys capture specific types of mobility that are not always included (for example, e-scooters, electric bikes and bike-sharing modes). Therefore, the precise numbers reported should be considered with caution. However, although the data might be imprecise, the principle is to capture that there are cities where most commuting is by $A$, like Utrecht in the Netherlands, some where it is mainly by $B$, like Minsk in Belarus, and others where it is mostly by $C$, like Charleroi in Belgium, Limerick in Ireland, or Palermo in Italy. The compiled data is open access and maintained at \href{https://citiesmoving.com/}{www.CitiesMoving.com}. 
}

%%%% filter europe
{
The data is not available for most cities worldwide, but there are nearly 400 observations in Europe (details in the Supplementary Material SM-A). Additionally, Europe has nearly 60\% of the length of the global tram network and generates almost 75\% of the total ridership \cite{UITP2023}, so it is possible to compare the share of $C$ between cities with a metro (47 cities), only a tram (46 cities), or none (285 cities). For city $i$, the modal share is represented by the $(A_i, B_i, C_i)$ triplet, so $A_i + B_i + C_i =100\%$ (Figure \ref{Figure1}). 

\begin{figure}[h!] \centering
\includegraphics[width=0.85\textwidth]{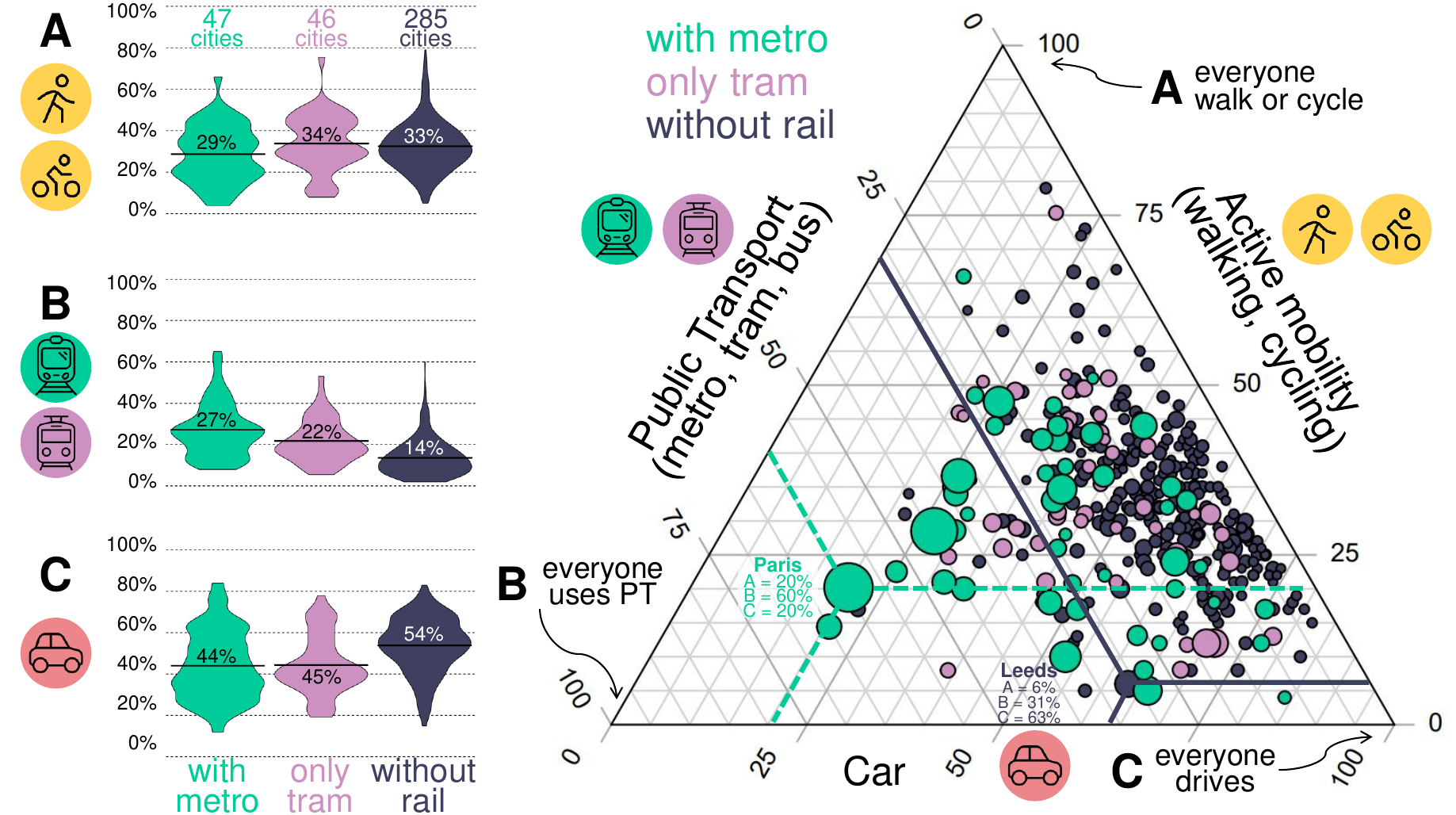}
\caption{$ABC$ framework for 378 cities in Europe. Left - Distribution of the modal share depending on whether they have a metro, only a tram, or no rail system. Right - $ABC$ for the 378 cities in Europe \cite{MobilityABCPrietoOspina}.}
\label{Figure1}
\end{figure}
}

%%%% there are differences. violins
{
To detect differences in the modal share of cities, they are grouped according to whether they have a metro, only a tram, or none. The number represents the weighted average of each group, where each observation is weighted by the population of its respective city (details in the Methods). For cities with a metro, the $A$ share is 0.275, the $B$ share is 0.352, and the $C$ share is 0.373 (Figure \ref{Figure2}). For cities with only a tram, the $A$ share is 0.286, the $B$ is 0.211, and the $C$ is 0.502. Finally, for cities without metro or tram, the $A$ share is 0.306, the $B$ is 0.157, and the $C$ is 0.537. Having a metro or tram system in the city is correlated with significant differences in terms of the modal share (details of the statistical significance in the SM-B). The share of $B$ is 0.135 higher if the city has a metro than if the city has no rail (with a p-value of $9.2 \times 10^{-9}$) and also 0.053 higher if the city has a metro than if the city has only a tram (with a p-value of 0.0153). Although cities with a metro tend to have a higher share of $B$, it is unclear whether the presence itself causes people to shift to $B$, and no causal claims can be made. At the city level, a combination of attributes, such as urban form and the convenience of car use, has been found to contribute significantly to levels of car dependency \cite{saeidizand2024patterns}. At the individual level, travel behaviour is influenced by various personal factors, such as age, income, attitudes, and residential self-selection, which shape preferences for car use and ownership \cite{saeidizand2022revisiting}. For instance, a person who prefers travelling by $B$ may choose to live in an area with better access to that mode \cite{schwanen2005affects}.

\begin{figure}[h!] \centering
\includegraphics[width=0.85\textwidth]{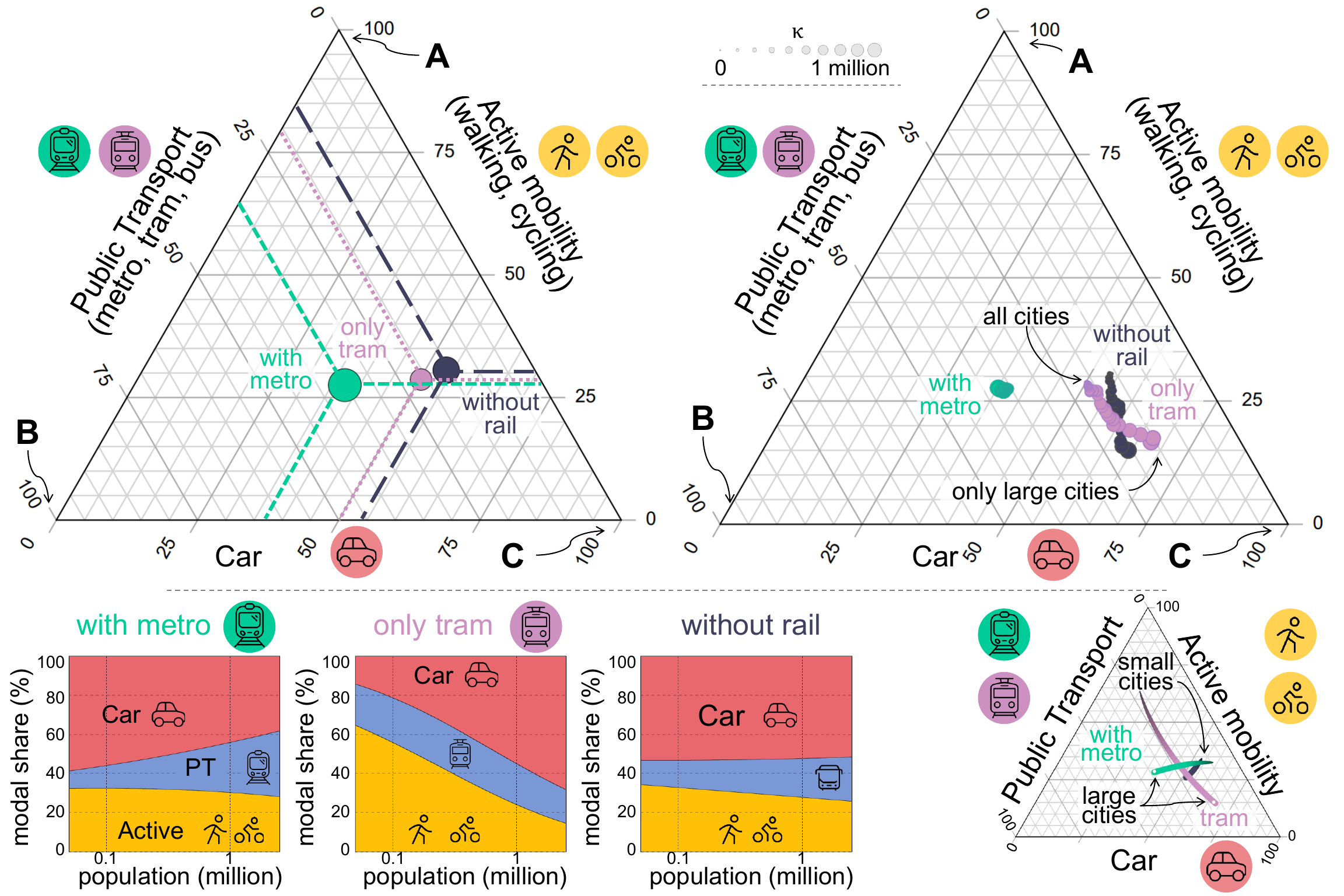}
\caption{$ABC$ framework depending on whether cities have a metro, only a tram or none. Left - Average $ABC$ share of cities in Europe with a metro, only a tram or none. The number represents the weighted average of each group, where each city is weighted by its population. Right - $ABC$ share when only cities above $\kappa$ size are considered. Bottom - $ABC$ share for cities with metro, only tram or neither, depending on the size of the population obtained with a Dirichlet regression model for compositional data, using \cite{maier2014dirichletreg, citeR}. The model displays the impact of whether the city has a metro, only a tram, or neither, as well as the population.  }
\label{Figure2}
\end{figure}
}

%%%% vary by city size. 
{
Cities vary considerably in population size, ranging from small urban agglomerations with a few thousand inhabitants to large metropolitan areas with millions of people, such as Paris or London. Population size is linked with modal share in three ways. First, with more people, commuting distances tend to be much longer, making Active Mobility more challenging. Second, with longer distances, driving is also more time-consuming, so using $C$ could be more inconvenient \cite{MacroTraffic25Cities}. However, large cities also tend to have better Public Transport provisions. Therefore, the observed drop in the $C$ share in cities with a metro could be an effect of city size. Considering only cities that have a population larger than some number, $\kappa$, it is possible to detect the differences in modal share across cities of varying sizes. For a group of cities (for example, those with a metro with more than half a million people), we compute the weighted modal share (where the weights correspond to the population of each city). We progressively increase the value of $\kappa$, so we first compute the modal share of those 47 cities with a metro, and then it progressively corresponds to the modal share of the 38 cities with more than half a million inhabitants, followed by the 27 cities with more than one million and so on. The process is an iterative filter that removes small cities from the analysis. Considering values of $\kappa$ from 0 (all cities) to 1 million people (so only 31 cities are considered, 25 of them with a metro, four with only a tram and two without a metro or tram) shows that modal share variations correspond only to cities without a metro (Figure \ref{Figure2}). The average modal share of cities with a metro remains the same regardless of whether only small or large cities are observed. In cities with a metro, the share of $A$ is 27\%, the share of $B$ is 37\%, and the share of $C$ is 36\%, independent of the value of $\kappa$. Thus, the longer distances experienced in bigger metropolitan areas do not necessarily result in a smaller share of $A$ if the city has a metro. However, this trend does not hold for cities without a metro, even those with a tram. For example, considering cities with only a tram, the share of $A$ is 29\%. However, for cities with more than 500,000 inhabitants, it drops to 22\%, and further decreases to 19\% for cities with more than 750,000 inhabitants (Figure \ref{Figure2}).
}

%%%% public transport metro, only tram, car
{
The share of $B$ remains roughly the same for cities with a metro (approximately 35\%), regardless of their size. However, for cities with only a tram, the share of $B$ is lower (approximately 20\%) and drops to 15\% if only large cities are considered (Figure \ref{Figure2}). Thus, for cities that have only a tram, city size is negatively correlated with the share of $B$. The opposite occurs for cities without rail, where the modal share of $B$ is approximately 15\% when all cities are considered, but it increases to 20\% when only large cities are taken into account.
}

%%%% car journeys yyy
{
There are also significant differences in the modal share of $C$ depending on whether the city has a metro, only a tram or none. In cities with a metro, the share of $C$ is 0.129 lower compared to cities that only have a tram and 0.164 lower compared to cities without a rail. Additionally, those differences tend to become more pronounced with city size. For example, considering only cities with more than 750,000 inhabitants, the share of $C$ is 36\% if they have a metro, but it is 63\% if they do not have a metro, regardless of whether they have a tram or not (Figure \ref{Figure2}). 
}

%%%% multinom
{
The correlation between the share of different modes of transport and the presence of various types of Public Transport can be modelled as follows. The modal share $Y_i = (A_i, B_i, C_i)$ is a triplet of proportions such that $A_i + B_i + C_i = 1$ for each city $i$ and each city is also characterised by a set of covariates $X_i$, including the presence of a metro or tram system and its size. A Dirichlet regression is designed for compositional data (where the dependent variables are proportions summing to one), accounting for the dependency between components \cite{maier2014dirichletreg}. The obtained coefficients for the Dirichlet regression, where the variables considered are dummy variables indicating whether the city has a metro, a tram, or neither, and the population (transformed on a logarithmic scale to reduce skewness), are presented in the Methods section. The results of the regression show that for cities with a metro, as the population increases, they tend to have a slightly smaller share of $C$ (Figure \ref{Figure2}-bottom). However, the opposite is observed in cities that only have a tram, similar to those without a railway system (see the Methods section for details).
}

%%% compare differences
{
To capture differences in modal share based on whether a city has a metro, only a tram, or neither, we compare the impact of population size. Across all cities with a metro, the share of $C$ is 15\% lower than in other cities (Figure \ref{Figure3}). However, this difference widens as the population size increases. For cities with over one million inhabitants, the gap in $C$ share between metro and non-metro cities reaches nearly 35\%. Additionally, the shares of $A$ and $B$ are higher in large cities with a metro. In cities without a metro, the share of $A$ exceeds 30\%, but it drops to less than 15\% in larger cities, most of this shift transferring to $C$. These differences are not observed in cities with only a tram. In small tram-only cities, the share of $B$ is slightly higher than in cities without one. However, in cities with over 800,000 inhabitants, the share of $B$ is lower in tram cities than in those without rail, with most of this shift transferring to $C$ (Figure \ref{Figure3}). Results show that when small cities are filtered, differences in the modal share become even more pronounced, indicating that the availability of efficient Public Transport, such as a metro, is vital in larger cities. 

\begin{figure}[h!] \centering
\includegraphics[width=0.85\textwidth]{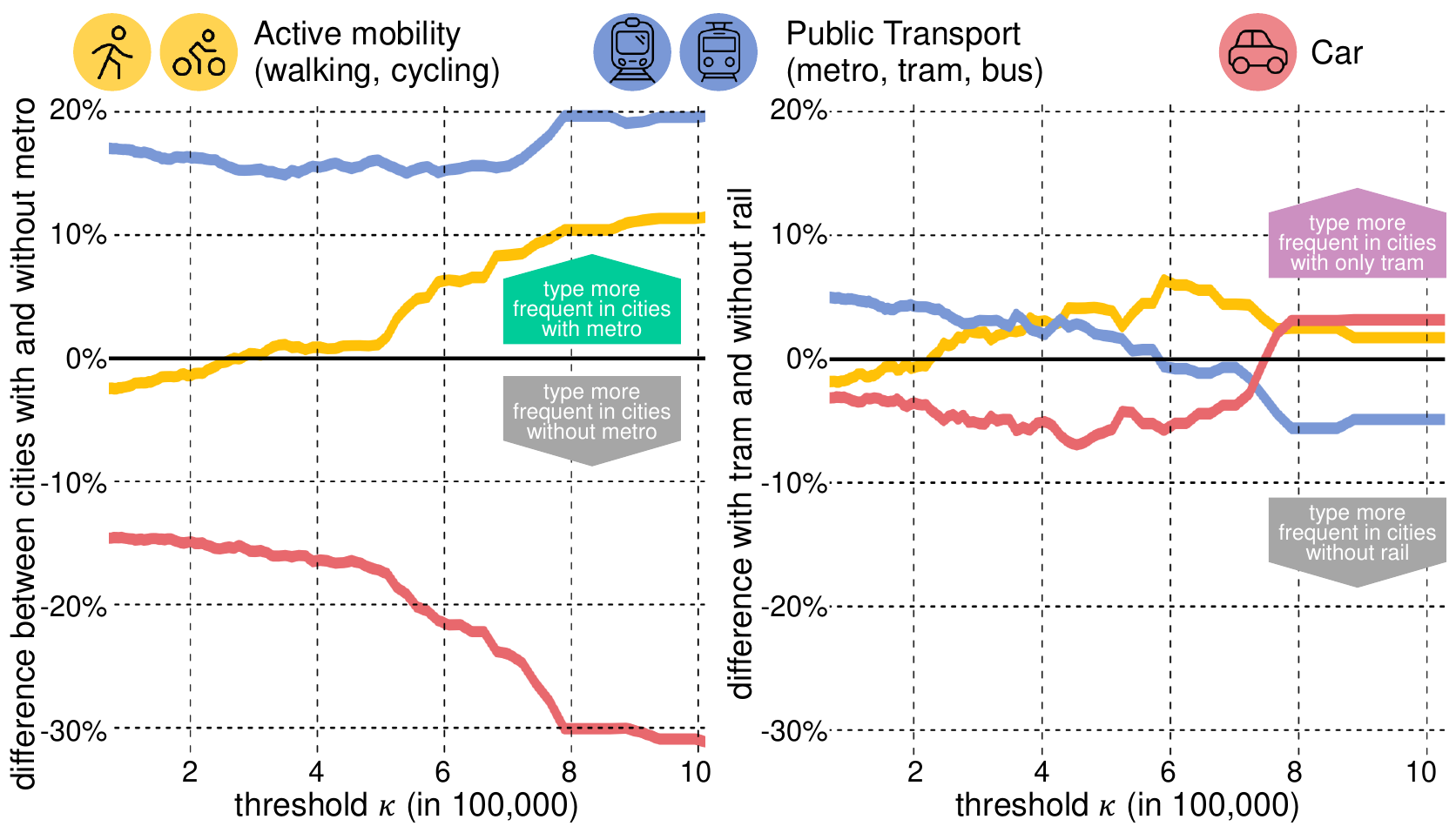}
\caption{Differences observed in the modal share across cities. Left - Differences for $ABC$ (vertical axis) between cities with a metro and cities without a metro (including cities with a tram), depending on the values of $\kappa$ considered (horizontal axis). Right - Differences for $ABC$ between cities with only a tram and cities without a metro or a tram. }
\label{Figure3}
\end{figure}
}

%%% Compare two cities of one million yy
{
To quantify the differences between cities with a metro, only a tram, or neither, we model the number of commutes by $C$ using the corresponding modal share. Consider three hypothetical cities, each with a population of one million: one with a metro, one with only a tram, and one with neither. Since the cities are identical in size, any differences in the number of $C$ journeys result solely from the presence or absence of a metro or tram system. Weekly commutes in each city are estimated using a two-step process. First, the total number of trips per person follows a fixed-rate distribution, capturing variability in individual travel behaviour. Second, each trip is assigned to one of the transport modes $A$, $B$, or $C$, based on the city's modal share, modelled as a Multinomial distribution. The total number of journeys for each mode is then aggregated (details are provided in the Methods section). Taking a city with one million inhabitants and the corresponding modal share, we can estimate the number of journeys in that city. If it has a metro System, the share of $C$ is 0.352, resulting in roughly 0.366 billion $C$ journeys in a year (Table \ref{OneMillionTable}). In a city with only a tram, the share of $C$ is 0.675, resulting in 0.702 billion $C$ journeys, 92\% more than in a city with a metro. Similarly, for a city with no metro or tram, the $C$ share is 0.663, resulting in 0.691 billion $C$ journeys per year.

\begin{table}[ht]
    \centering
    \begin{tabular}{rccc}
billion journeys & $A$     & $B$ & $C$ \\
\hline
with a metro & 0.288 & 0.385 & 0.366\\
only tram & 0.173 & 0.163 & 0.702\\
without metro or tram & 0.167 & 0.182 & 0.691\\
    \end{tabular}
    \caption{Estimated yearly journeys by $A$, $B$ and $C$ in a city with one million inhabitants, depending on whether it has a metro, only a tram or neither. }
    \label{OneMillionTable}
\end{table}
}

%%% close result
{
These results highlight the significant impact of transit infrastructure on modal share. Large cities with a metro have roughly half as many $C$ journeys as those without one. Also, even small reductions in the share of $C$ journeys can lead to substantial collective benefits. In a city like Vienna, for example, the modal share is $A = 34\%$, $B = 39\%$, and $C = 27\%$. Reducing the share of $C$ journeys from 27\% to 26\% and reallocating those trips to $A$ or $B$ would result in a nearly 4\% decrease in total car journeys—equivalent to almost six million fewer car trips per year.
}
\section{Discussion}

%%%% ABC first Metro reduces the number of journeys but Tram does not y
{
The $ABC$ framework of modal share across European cities was used to analyse whether the presence of a metro, a tram, or neither correlates with a higher share of $C$ journeys. Grouping different transport modes simplifies the analysis, as urban mobility encompasses various options, including private vehicles, motorbikes, taxis, and mobility apps. Using the $ABC$ framework, it was found that cities with a metro generally have a lower share of $C$, whereas this pattern does not hold consistently for cities with only a tram. 
}

%%%% metro yes
{
Metro systems are associated with a reduced $C$ share. However, some relevant exceptions exist. Many cities without a metro maintain a low share of $C$ but compensate their travel demands with a high share of $A$, such as Utrecht (Netherlands), Szeged (Hungary), and Bern (Switzerland), where $C$ represents only 20\% of the modal share. Conversely, some cities with a metro still exhibit a high $C$ share, exceeding two-thirds of the total, as seen in Charleroi (Belgium), Malaga (Spain), Toulouse (France), Brescia (Italy), and Rome (Italy).
}

%%%% on Active mobility TO CHECK!!
{
Active Mobility is the most sustainable form of transport \cite{de2019satisfying}. However, walking and cycling become less practical in large cities due to greater distances, frequently leading to a decline in their share as city size increases \cite{MobilityABCPrietoOspina}. However, the results show that in cities with a metro, the share of $A$ remains stable (at around 30\%), regardless of city size. This trend does not hold for cities without a metro, where the share of $A$ decreases rapidly for large cities, even if they have a tram. This pattern likely reflects many underlying factors, such as car ownership, that were not taken into account. Cities with a metro may also be more receptive to policies that reduce car use and promote alternative mobility, as the metro provides a viable commuting alternative. However, these findings are correlational, and no direct causal relationship can be established. 
}

%%%%% issues with our metric
{
Many factors beyond Public Transport provisions can significantly influence car usage, including urban form (with elements such as density, polycentrism, and fragmentation) and financial considerations (such as fuel prices and Public Transport fares), all of which vary considerably between cities. Moreover, substantial differences exist between metro systems (some of which are rudimentary rail systems with low coverage and frequency), tram systems, and some bus-based systems, which are classified as high-quality. Thus, treating their presence as a simple binary variable at the city level is limiting. This approach overlooks critical indicators of Public Transport quality, such as the number and spatial distribution of stations, travel times, and service reliability \cite{Tram19Cities}. These elements could help understand why in some cities with a metro, the share of $C$ is high. For example, Toulouse (France), Palma de Mallorca (Spain), and Catania (Italy) are medium-sized cities with a metro with a share of $C$ close to 70\%. In the case of Palma or Catania, for example, the metro only covers a limited part of the city, so that network does not meet most travel demands. Similarly, the tram was only considered to be a covariate in cities without a metro, so it operates as a replacement for the metro network. However, in many cities with a metro, the tram serves as an intermediate mode of transport, increasing connections and resilience within the public transport system \cite{Tram19Cities}. 
}

%%%% cars and sustainable mobility y
{
Reducing car ownership and usage is essential for sustainable cities. The negative impact of cars extends way beyond fuel consumption, including noise, road accidents, pollution, urban heat, infrastructure and tyre wear. Encouraging fewer and shorter trips, promoting Active Mobility for short distances, and prioritising Public Transport for longer journeys are key strategies for sustainable urban living. Cities have implemented various policies to reduce car dependence, such as limiting parking, reallocating car lanes for cycling infrastructure, and expanding green infrastructure like bike lanes and pedestrian zones \cite{lindsey2023distributional}. These efforts are often coupled with promoting compact, mixed-use development to shorten travel distances \cite{MacroTraffic25Cities}. Among the key strategies for achieving more sustainable urban environments, Public Transport plays a crucial role in shaping mobility patterns \cite{saeidizand2025car}. Improved Public Transport provisions support compact and efficient urban growth \cite{GASCON2020362, ewing2017does}. However, Public Transport systems are often fragmented and insufficient, with many users being ``captive riders'' who lack viable alternatives \cite{oviedo2022accessibility}. In particular, poor households in peripheral areas frequently experience the lowest quality of service, characterised by limited speed, frequency, affordability, and reliability \cite{hernandez2016transport}. Enhancing Public Transport services not only improves overall mobility but also benefits marginalised communities, including low-income residents, youth, the elderly, and women, ultimately contributing to greater equity and quality of life \cite{macedo2022differences}.
}

\clearpage

\section{Methods}

\subsection{Modal share data}

%%%% introduce ABC
{
The data corresponding to the modal share in cities was obtained by combining mobility surveys conducted at the city level \cite{MobilityABCPrietoOspina}. Different sources were employed, including the European Platform on Mobility Management (EPOMM), Urban Audit, and Local Governments for Sustainability (ICLEI), among others \cite{epomm, UE_urbanaudit, ICLEI}. The data gives an estimate of the share of different modes of transport for regular commutes. The data was then aggregated into different categories, including all modes of Active Mobility ($A$), all modes of Public Transport ($B$), and private vehicles ($C$), which encompassed cars, motorbikes, taxis, and mobility platforms. For city $i$ the modal share is represented by $A_i$, $B_i$ or $C_i$, where $A_i + B_i + C_i = 100\%$. The data is open-access, updated with recent observations, and available at \href{https://citiesmoving.com/}{www.CitiesMoving.com}, including observations outside of Europe. 
}

%%%% why Europe
{
Although the data also includes observations from more than 600 cities outside Europe, a similar analysis is not possible in other regions. Many observations correspond to cities in the US, where mobility is dominated by cars, regardless of whether or not they have Public Transport provisions \cite{MobilityABCPrietoOspina}. Further, observations from Latin America, Asia or Africa are mainly from large cities, most of them with a metro (like Mexico City, Tokyo, or Buenos Aires), so they do not form a representative sample of cities. Thus, the analysis in Europe is ideal for analysing the use of Cars and Public Transport provisions. 
}

%%%% Metro and Tram data
{
Detecting whether cities have a metro, a tram, or none was done by an extensive search for Public Transport options in each city. Having a metro or a tram was classified as a binary variable, so attributes such as the number of stations, lines, km covered, cost per journey or the frequency of services are not considered \cite{vanoutrive2015modal, Tram19Cities}. Further, the classification does not include cities with train stations used for interurban mobility. 
}

\subsection{Combined frequency of a modal share} %%%% here explain also a threshold with \kappa and differences

%%%% combine ABC for cities of different size
{
Different groups of cities are analysed, including those with a metro, those that have a tram but not a metro and those that have neither. For a group of cities, $G$, the share of $A$ is computed as
\begin{equation}
A_G = \frac{\sum_{i \in G} A_i P_i}{\sum_{i \in G} P_i },
\end{equation}
where $P_i$ corresponds to the population of the city $i$. For a group of cities, $A_G$ is the average share of Active Mobility of each city, weighted by its population. Similarly, $B_G$ and $C_G$ correspond to the weighted share of $B$ and $C$.
}

%%%% kappa
{
A metro system is rarely observed in small cities. To control the effect of city size, groups of cities are formed by considering their population size. For some value of $\kappa > 0$, the group $G(\kappa)$ is formed by cities where $P_i > \kappa$. Then, the share of $A$ for those cities, $A_{G(\kappa) }$, is computed. For two groups of cities $G$ and $G'$, it is possible to compare their share $A_G - A_{G'}$, and similarly for other modes of transport. The principle here is to filter cities above a certain size and detect if the differences in the modal share between those with a metro, only a tram or neither are observed. 
}

\subsection{Modelling the modal share in cities}

%%%intro to model
{
The modal share $Y_i = (A_i, B_i, C_i)$ is a triplet of proportions such that $A_i + B_i + C_i = 1$ for each city $i$, representing the share of journeys made by different transport modes. Each city is also characterised by a set of covariates $X_i$, including the presence of a metro, a tram, and city size. There are several ways to model $Y_i$ as a function of $X_i$ using a regression technique. There are, however, two critical elements to consider before applying a regression. The variables $A_i, B_i, C_i$ are bounded in the $[0,1]$ interval, so a linear regression directly to the observations may lead to predictions outside bounds, heteroskedastic errors, and bias. One option is to perform a variable transformation (for example, by considering $A_i^\star = \log (A_i / (1-A_i))$ where the variable $A_i^\star$ is not bounded). Transformations convert the data into unconstrained values, allowing standard regression techniques to be applied. However, this approach can be problematic when some components are close to zero, as is the case for $B_i$ in some cities, and we would be ignoring that $A_i + B_i + C_i = 1$. In such cases, Dirichlet regression and multinomial logistic regression may be more appropriate. 
}

%%% DIRICHLET
{
A Dirichlet regression is a statistical model that is specifically designed for compositional data (when the outcome variables are proportions summing to one). It accounts for the dependency between components and models how explanatory variables influence the relative composition of the proportions \cite{maier2014dirichletreg}. The model respects both the $[0,1]$ bounds and the constraint of $A_i + B_i + C_i = 1$ inherent to compositional data. We specify a Dirichlet regression model in which the response vector $Y$, representing compositional data, is expressed as a function of the predictor $X$, considering size and the presence of transport infrastructure. The Dirichlet regression is computed using \cite{maier2014dirichletreg, citeR}. The variables considered in the regression are dummy variables that capture whether the city has a metro, a tram, or none (with names $HasMetro$, $HasTram$, and $HasRail$), and the population, which was log-transformed to account for scale effects and reduce skewness. Terms like $\text{log(pop)}:HasMetro$' are interaction terms, which are included to capture whether the effect of population size may differ depending on transit infrastructure. 
}

%%%%% results dirichlet
{
To interpret the coefficients, we compute the modelled share of people commuting by $A$ directly as
\begin{equation} \label{predict}
\frac{\exp{(X \beta_A)}}{\exp{(X \beta_A)}+\exp{(X \beta_B)}+\exp{(X \beta_C)}},
\end{equation}
where $X$ corresponds to the variable values being modelled (for example, if the city has a metro, the dummy variables $HasMetro$ and $HasRail$ are both equal to one, as well as its log population), and the values of $\beta_A$, $\beta_B$ and $\beta_C$ are the coefficients reported. All coefficients in the SM-C. 
}

%%%% multinomial
{
Interpreting the coefficients obtained is not straightforward, as they include interaction terms and the ratio of different modes of transport from Equation \ref{predict}. Therefore, even though the coefficient for $HasMetro$ is $+13.3239$ for $C$ journeys, this does not imply that cities with a metro have more car journeys. The coefficients obtained from the regression reflect relative effects within a compositional framework. There are several reasons for this. First, cities with a metro system also have a rail system, meaning the $HasRail$ variable is also active. Second, there is an interaction term with the (log) population. Additionally, in a Dirichlet regression, the modal share for each mode must be interpreted relative to the others (as shown in Equation \ref{predict}). Thus, the most direct way to assess the impact of specific variables, such as having a metro, a tram, or neither, is to plot the predicted modal shares across a range of population sizes for different infrastructure configurations. Figure 3 in the manuscript illustrates the effects of these variables on modal share, depending on whether a city has a metro, a tram, or neither, as well as city size. The results from the Dirichlet regression show that the share of $A$ decreases with increasing population, but the presence of a metro system boosts it. Conversely, the presence of only a tram system is associated with a significant reduction in the share of $A$, particularly in larger cities where the share of $C$ is considerably higher.
}

\subsection{Estimating the number of car journeys in a city}

%%% take pois then multinomial then...
{
The number of weekly $C$ commutes in each city is estimated in two steps. First, for each person in city $i$, the number of weekly journeys is assumed to be drawn from a distribution with a mean rate of $\mu$. Each person is expected to take $\mu$ trips per week, with some variability in the actual number of journeys. This variability reflects differences in travel behaviour among individuals and fluctuations from week to week. A Poisson distribution with a homogeneous rate of $\mu = 20$ trips per week is used. While other distributions, such as the Negative Binomial or Geometric distribution, could be applied for trip generation, the Poisson distribution is commonly used to model trip frequency \cite{barmby1989modelling, mukherjee2022comprehensive, chang2014comparative, jang2005count, kim2013comparison}. A baseline rate of approximately 20 trips per week, based on survey data, is used to simulate journeys \cite{UK_2020}. The actual number of weekly trips may vary based on factors such as age and income. For instance, in the US, it was observed that for every additional U\$100 in income, a household takes approximately one extra trip per year \cite{US_2017}. However, to maintain simplicity in the trip generation process, a constant rate is assumed.
}

{
The second step is to decide whether each journey is by $A$, $B$, or $C$. The choice of transport mode is modelled using the modal share of city $i$ as a Multinomial distribution, with probabilities $\pi_i = (A_i, B_i, C_i)$. While transport mode choice varies across individuals, the $ABC$ data does not capture this level of granularity, so a homogeneous modal share is assumed for the population. The number of journeys a person takes by $A$, $B$ or $C$ is considered a Multinomial distribution. Additionally, since a Multinomial distribution conditional on a Poisson-distributed total (the number of weekly journeys) also follows a Poisson distribution, the number of $A$ journeys follows $\text{Pois}(\mu A_i)$. The total number of journeys for the entire city is the sum of individual journeys. Since the sum of independent Poisson-distributed variables also follows a Poisson distribution, the number of $A$ journeys in city $i$ during a typical week follows $\text{Pois}(\mu A_i P_i)$. Therefore, the expected number of $A$ journeys in city $i$ is $\mu A_i P_i$, with a standard deviation of $\sqrt{\mu A_i P_i}$. The same is done for the number of $B$ journeys and the number of $C$ journeys.
}

\section{Data availability}

The data used for this study is open-access, available at \href{https://citiesmoving.com/}{www.CitiesMoving.com}.

\section*{Acknowledgements}

The research was funded by the Austrian Federal Ministry for Climate Action, Environment, Energy, Mobility, Innovation and Technology (2021-0.664.668) and the Austrian Federal Ministry of the Interior (2022-0.392.231).

\section*{Author's contributions}

The author confirms sole responsibility for the conception of the study, the presentation of results, and manuscript preparation.

\section*{Competing interests}

The author declares having no competing interests.

\clearpage

\section*{Supplementary information}

\subsection*{A - Mobility data across cities}

%%% data was compiled and made accessible
{
The mobility data concentrates on metropolitan areas, gathered from many sources, including the European Platform on Mobility Management (EPOMM), and ICLEI, with data supplied by the EcoMobility Alliance Cities \cite{Wiersma2021, Duleba2021, GovernmentOfficeforScience2018}. The compiled data is available and maintained at www.CitiesMoving.com. Given the diversity of the available transportation modes for all cities, the modal share is simplified into the $ABC$ scheme, which differentiates Active Mobility ($A$), Public Transport ($B$) and Cars ($C$). For city $i$, the proportion of all types of Active Mobility journeys (including walking, cycling and others) is merged into a single category $A_i$. The share of all Public Transport journeys (including metro, tram, BRT, and others) is incorporated into a category, $B_i$, and all private motorised journeys (including cars, taxis, and others) into $C_i$. For each category, its relative frequency is considered, ensuring that $A_i + B_i + C_i = 100\%$, and it is displayed on a ternary plot.
}

%%%%
{
Comparing modal share across cities is challenging since it depends on the definition of the urban area, the type of journeys, the different modes of transport, and how multi-modal journeys are reported. Here, the data indicate the primary mode of transportation for each trip \cite{MobilityABCPrietoOspina}.
}

%%% Metro and Tram data was also compiled
{
Detecting whether cities have a metro, a tram, or none was done by an extensive search for Public Transport options in each city. The year of construction of the lines, the number of stations, passengers or other details about the metro or tram lines were not considered in the analysis.
}

%%%% why Europe
{
Mobility across cities in Europe has been analysed for two reasons. First is that the $ABC$ modal share data is available for nearly 400 cities \cite{MobilityABCPrietoOspina}. In other regions, mobility data is scarce and biased towards big cities, so most of them have a functioning metro system. In the US, the high levels of car dependency make it less ideal to detect the impact of Public Transport. Second, Europe is the region with the most well-established and widespread network of trams. Europe has nearly 60\% of the km of the global network and generates almost 75\% of the total ridership \cite{UITP2023}, but there are also many cities that have a metro system or that have a tram but no metro. Across Europe, only 12\% of cities are found to have a metro, and another 13\% of cities are found to have only a tram.
}

\subsection*{B - Significant differences in modal share across cities with metro, tram or none}

%%%% difference in mean
{
The weighted average of cities with a metro, with only a tram or with no rail for the $ABC$ is computed by considering the population of each observation. For example, if city $A$ has one million people and city $B$ has ten million, then for the computation, $B$ is assigned ten times more weight in the calculation. The weighted average can be interpreted as the modal share of a randomly selected person from the group of cities. It is obtained that for cities with a metro, the $A$ share is 0.275, the $B$ share is 0.352, and the $C$ share is 0.373. For cities with only a tram, the $A$ share is 0.286, the $B$ is 0.211, and the $C$ is 0.502. Finally, for cities without metro or tram, the $A$ share is 0.306, the $B$ is 0.157, and the $C$ is 0.537. 
}

{
It is possible to test whether the share of people using a Car is statistically different depending on whether the city has a metro, a tram or none. An unpaired t-test is used to check whether there is any statistical difference in the mean between the groups. There is no statistical difference in the share of Active Mobility between cities that have a metro, those that have only a tram, or those that have neither. There is a difference in the number of Public Transport journeys between cities with a metro (mean of 0.271) and cities without a metro or tram (mean of 0.136). The share of Public Transport, thus, is 0.135 higher if the city has a metro (with a p-value of $9.2 \times 10^{-9}$). Also, there is a difference in the number of Public Transport journeys in cities that have a metro and cities that only have a tram (mean of 0.217), with a difference in the share of 0.053 (with a p-value of 0.0153). These differences are calculated without considering weights for each city. The Car share in cities with a metro is 0.44, and the Car share in a city with no metro or tram is 0.538. There is a statistical difference between them of 0.098 (with a p-value of $1.8 \times 10^{-4}$). There is also a statistical difference in the Car share if the city has a tram (with a mean of 0.444), with a difference of 0.094 (with a p-value of $1.1 \times 10^{-4}$).
}

\subsection*{C - Regression coefficients for transport mode shares}

Supplementary Table \ref{DirichletCoefficients} reports the coefficients obtained in the Dirichlet regression model, where the response vector $Y$, representing compositional data, is expressed as a function of the predictor $X$, considering size and the presence of transport infrastructure.

\begin{table}[ht]
\centering
\begin{tabular}{lrrrr}
\multicolumn{5}{c}{Dirichlet regression coefficients for transport mode shares} \\
\hline
\textbf{Variable} & \textbf{Estimate} & \textbf{Std. Error} & \textbf{z value} & \textbf{Pr($>|z|$)} \\
\hline
\multicolumn{5}{c}{Component: $A$ Active Mobility } \\
(Intercept) & 3.0343 & 0.0390 & 77.71 & $<\!2\text{e-}16^{***}$ \\
log(population) & -0.1215 & 0.0032 & -37.49 & $<\!2\text{e-}16^{***}$ \\
HasMetro & 2.0179 & 0.1871 & 10.78 & $<\!2\text{e-}16^{***}$ \\
HasTram & -2.3638 & 0.1986 & -11.90 & $<\!2\text{e-}16^{***}$ \\
HasRail & -5.0880 & 0.2550 & -19.95 & $<\!2\text{e-}16^{***}$ \\
log(pop):HasMetro & -0.1966 & 0.0138 & -14.26 & $<\!2\text{e-}16^{***}$ \\
log(pop):HasTram & 0.1836 & 0.0145 & 12.69 & $<\!2\text{e-}16^{***}$ \\
log(pop):HasRail & 0.3927 & 0.0189 & 20.76 & $<\!2\text{e-}16^{***}$ \\
\hline
\multicolumn{5}{c}{Component: $B$ Public Transport} \\
(Intercept) & -0.4195 & 0.0421 & -9.97 & $<\!2\text{e-}16^{***}$ \\
log(population) & 0.1038 & 0.0035 & 29.65 & $<\!2\text{e-}16^{***}$ \\
HasMetro & 1.3269 & 0.2085 & 6.36 & $1.97\text{e-}10^{***}$ \\
HasTram & -4.2627 & 0.2271 & -18.77 & $<\!2\text{e-}16^{***}$ \\
HasRail & -4.3977 & 0.2894 & -15.20 & $<\!2\text{e-}16^{***}$ \\
log(pop):HasMetro & -0.1485 & 0.0153 & -9.67 & $<\!2\text{e-}16^{***}$ \\
log(pop):HasTram & 0.3042 & 0.0166 & 18.37 & $<\!2\text{e-}16^{***}$ \\
log(pop):HasRail & 0.3739 & 0.0214 & 17.43 & $<\!2\text{e-}16^{***}$ \\
\hline
\multicolumn{5}{c}{Component: $C$ Cars} \\
(Intercept) & 2.7708 & 0.0480 & 57.76 & $<\!2\text{e-}16^{***}$ \\
log(population) & -0.0569 & 0.0040 & -14.25 & $<\!2\text{e-}16^{***}$ \\
HasMetro & 13.3239 & 0.2040 & 65.32 & $<\!2\text{e-}16^{***}$ \\
HasTram & -7.3349 & 0.1620 & -45.29 & $<\!2\text{e-}16^{***}$ \\
HasRail & -9.7666 & 0.2507 & -38.96 & $<\!2\text{e-}16^{***}$ \\
log(pop):HasMetro & -1.0476 & 0.0152 & -69.11 & $<\!2\text{e-}16^{***}$ \\
log(pop):HasTram & 0.5302 & 0.0118 & 44.99 & $<\!2\text{e-}16^{***}$ \\
log(pop):HasRail & 0.7587 & 0.0187 & 40.57 & $<\!2\text{e-}16^{***}$ \\
\hline
\end{tabular}
\vspace{1ex}
\caption{All coefficients are from a Dirichlet regression model using a log link function with a common parametrisation. They were obtained using \cite{maier2014dirichletreg, citeR}. In all cases, the null hypothesis is that the coefficient is zero, and it is rejected by a one-sided test. The significance codes for the p-values are $^{***}p < 0.001$.} 
\label{DirichletCoefficients}
\end{table}

\clearpage

\bibliographystyle{unsrt}

\begin{thebibliography}{10}

\bibitem{miner2024car}
Patrick {Miner}, Barbara~M {Smith}, Anant {Jani}, Geraldine {McNeill}, and Alfred {Gathorne-Hardy}.
\newblock Car harm: A global review of automobility's harm to people and the environment.
\newblock {\em Journal of Transport Geography}, 115:103817, 2024.

\bibitem{berrill2024comparing}
Peter {Berrill}, Florian {Nachtigall}, Aneeque {Javaid}, Nikola {Milojevic-Dupont}, Felix {Wagner}, and Felix {Creutzig}.
\newblock Comparing urban form influences on travel distance, car ownership, and mode choice.
\newblock {\em Transportation Research Part D: Transport and Environment}, 128:104087, 2024.

\bibitem{haberl2021stocks}
Helmut {Haberl}, Martin {Schmid}, Willi {Haas}, Dominik {Wiedenhofer}, Henrike {Rau}, and Verena {Winiwarter}.
\newblock Stocks, flows, services and practices: Nexus approaches to sustainable social metabolism.
\newblock {\em Ecological Economics}, 182:106949, 2021.

\bibitem{aiello2024urban}
Luca~Maria {Aiello}, Anastassia {Vybornova}, S{\'a}ndor {Juh{\'a}sz}, Michael {Szell}, and Eszter {Bok{\'a}nyi}.
\newblock Urban highways are barriers to social ties.
\newblock {\em Proceedings of the National Academy of Sciences}, 122(10):e2408937122, 2025.

\bibitem{cuenot2017wider}
Francois {Cuenot}.
\newblock Wider, taller, heavier: Evolution of light-duty vehicle size over generations.
\newblock {\em Working paper 17}, 2017.

\bibitem{yoshimura2022street}
Yuji {Yoshimura}, Yusuke {Kumakoshi}, Yichun {Fan}, Sebastiano {Milardo}, Hideki {Koizumi}, Paolo {Santi}, Juan~Murillo {Arias}, Siqi {Zheng}, and Carlo {Ratti}.
\newblock Street pedestrianization in urban districts: Economic impacts in {S}panish cities.
\newblock {\em Cities}, 120:103468, 2022.

\bibitem{PrietoInventarioMovilidad}
Rafael {Prieto-Curiel}.
\newblock El inventario de la movilidad de {M}éxico, 2023.

\bibitem{DoubleCars}
Matthew~Nitch {Smith}.
\newblock The number of cars worldwide is set to double by 2040, 2016.

\bibitem{grubler2003technology}
Arnulf {Gr{\"u}bler}.
\newblock {\em Technology and global change}.
\newblock Cambridge University Press, Cambridge, UK, 2003.

\bibitem{krueger2009national}
Alan~B {Krueger}, Daniel {Kahneman}, David {Schkade}, Norbert {Schwarz}, and Arthur~A {Stone}.
\newblock National time accounting: The currency of life.
\newblock In {\em Measuring the subjective well-being of nations: National accounts of time use and well-being}, pages 9--86. University of Chicago Press, Chicago, USA, 2009.

\bibitem{prieto2022ubiquitous}
Rafael {Prieto-Curiel}, Humberto {Gonz{\'a}lez Ram{\'\i}rez}, and Steven {Bishop}.
\newblock A ubiquitous collective tragedy in transport.
\newblock {\em Frontiers in Physics}, page 488, 2022.

\bibitem{van2010perceptions}
Nicolaas Jacob~Arnold {van Exel} and Piet {Rietveld}.
\newblock Perceptions of public transport travel time and their effect on choice-sets among car drivers.
\newblock {\em Journal of Transport and Land Use}, 2(3/4):75--86, 2010.

\bibitem{pisoni2022active}
Enrico {Pisoni}, Panayotis {Christidis}, and E~{Navajas Cawood}.
\newblock Active mobility versus motorized transport? user choices and benefits for the society.
\newblock {\em Science of The Total Environment}, 806:150627, 2022.

\bibitem{nieuwenhuijsen2020urban}
Mark~J {Nieuwenhuijsen}.
\newblock Urban and transport planning pathways to carbon neutral, liveable and healthy cities; a review of the current evidence.
\newblock {\em Environment International}, 140:105661, 2020.

\bibitem{croeser2022finding}
Thami {Croeser}, Georgia~E {Garrard}, Casey {Visintin}, Holly {Kirk}, Alessandro {Ossola}, Casey {Furlong}, Rebecca {Clements}, Andrew {Butt}, Elizabeth {Taylor}, and Sarah~A {Bekessy}.
\newblock Finding space for nature in cities: the considerable potential of redundant car parking.
\newblock {\em NPJ Urban Sustainability}, 2(1):1--13, 2022.

\bibitem{liao2020disparities}
Yuan {Liao}, Jorge {Gil}, Rafael~HM {Pereira}, Sonia {Yeh}, and Vilhelm {Verendel}.
\newblock Disparities in travel times between car and transit: Spatiotemporal patterns in cities.
\newblock {\em Scientific Reports}, 10(1):1--12, 2020.

\bibitem{banister2011cities}
David {Banister}.
\newblock Cities, mobility and climate change.
\newblock {\em Journal of Transport Geography}, 19(6):1538--1546, 2011.

\bibitem{de2016travel}
Jonas {De Vos}, Patricia~L {Mokhtarian}, Tim {Schwanen}, Veronique {Van Acker}, and Frank {Witlox}.
\newblock Travel mode choice and travel satisfaction: bridging the gap between decision utility and experienced utility.
\newblock {\em Transportation}, 43(5):771--796, 2016.

\bibitem{MultimodalSzell2022}
Laura {Alessandretti}, Luis~Guillermo {Natera Orozco}, Federico {Battiston}, Meead {Saberi}, and Michael {Szell}.
\newblock Multimodal urban mobility and multilayer transport networks.
\newblock {\em Environment and Planning B: Urban Analytics and City Science}, 1(1):23998083221108190, 2022.

\bibitem{douglas2011cars}
Margaret~J {Douglas}, Stephen~J {Watkins}, Dermot~R {Gorman}, and Martin {Higgins}.
\newblock Are cars the new tobacco?
\newblock {\em Journal of Public Health}, 33(2):160--169, 2011.

\bibitem{wang2024quantification}
Ziyulong Wang, Ketong Huang, Renzo Massobrio, Alessandro Bombelli, and Oded Cats.
\newblock Quantification and comparison of hierarchy in public transport networks.
\newblock {\em Physica A: Statistical Mechanics and its Applications}, 634:129479, 2024.

\bibitem{yang2017choosing}
Wenhan {Yang} and John {Zacharias}.
\newblock Choosing between {T}ram and {M}etro in {H}ong {K}ong--utility, affect and demographics.
\newblock {\em WIT Transactions on the Built Environment}, 176:131--141, 2017.

\bibitem{lin2022metro}
Dong {Lin}, Wout {Broere}, and Jianqiang {Cui}.
\newblock Metro systems and urban development: Impacts and implications.
\newblock {\em Tunnelling and Underground Space Technology}, 125:104509, 2022.

\bibitem{UITP2021}
International~Association of~Public Transport~(UITP).
\newblock World {M}etro {F}igures 2021.
\newblock Technical report, UITP, 2021.
\newblock Accessed: 2025-02-22.

\bibitem{UITP2023}
International~Association of~Public Transport~(UITP).
\newblock The {G}lobal {T}ram and {L}ight {R}ail {L}andscape 2019-21.
\newblock Technical report, UITP, 2023.
\newblock Accessed: 2025-02-22.

\bibitem{redman2013quality}
Lauren {Redman}, Margareta {Friman}, Tommy {G{\"a}rling}, and Terry {Hartig}.
\newblock Quality attributes of public transport that attract car users: A research review.
\newblock {\em Transport Policy}, 25:119--127, 2013.

\bibitem{beirao2007understanding}
Gabriela {Beir{\~a}o} and JA~Sarsfield {Cabral}.
\newblock Understanding attitudes towards public transport and private car: A qualitative study.
\newblock {\em Transport Policy}, 14(6):478--489, 2007.

\bibitem{reinhold2008more}
Tom {Reinhold}.
\newblock More passengers and reduced costs—the optimization of the {B}erlin public transport network.
\newblock {\em Journal of Public Transportation}, 11(3):4, 2008.

\bibitem{levinson2007planning}
David~M {Levinson} and Kevin~J {Krizek}.
\newblock {\em Planning for place and plexus: Metropolitan land use and transport}.
\newblock Routledge, New York, USA, 2007.

\bibitem{saeidizand2025car}
Pedram {Saeidizand}, Perseverence {Savieri}, and Kobe {Boussauw}.
\newblock Car dependency contributors in global metropolitan areas over time.
\newblock {\em Journal of Transport Geography}, 123:104152, 2025.

\bibitem{mogridge1997self}
Martin~JH {Mogridge}.
\newblock The self-defeating nature of urban road capacity policy: A review of theories, disputes and available evidence.
\newblock {\em Transport Policy}, 4(1):5--23, 1997.

\bibitem{bar2013model}
Asaf {Bar-Yosef}, Karel {Martens}, and Itzhak {Benenson}.
\newblock A model of the vicious cycle of a bus line.
\newblock {\em Transportation Research Part B: Methodological}, 54:37--50, 2013.

\bibitem{prieto2021paradox}
Rafael {Prieto-Curiel}, Humberto {Gonz{\'a}lez-Ram{\'\i}rez}, Mauricio {Qui{\~n}ones-Dom{\'\i}nguez}, and Juan~Pablo {Orjuela-Mendoza}.
\newblock A paradox of traffic and extra cars in a city as a collective behaviour.
\newblock {\em Royal Society Open Science}, 8(6):201808, 2021.

\bibitem{MobilityABCPrietoOspina}
Rafael {Prieto-Curiel} and Juan~P. {Ospina}.
\newblock The {ABC} of mobility.
\newblock {\em Environment International}, 185:108541, 2024.

\bibitem{Wiersma2021}
J.~K. {Wiersma}, L.~{Bertolini}, and L-{Harms}.
\newblock Spatial conditions for car dependency in mid-sized {E}uropean city regions.
\newblock {\em European Planning Studies}, 29(7):1314--1330, 2021.

\bibitem{Duleba2021}
Szabolcs Duleba, Sarbast Moslem, and Domokos Eszterg{\'{a}}r-Kiss.
\newblock Estimating commuting modal split by using the best-worst method.
\newblock {\em European Transport Research Review}, 13(1), 2021.

\bibitem{GovernmentOfficeforScience2018}
{Government Office for Science}, {University of Birmingham}, and Miles Tight.
\newblock {Walking in the UK transport system: how and why is it changing?}, 2018.

\bibitem{saeidizand2022revisiting}
Pedram {Saeidizand}, Koos {Fransen}, and Kobe {Boussauw}.
\newblock Revisiting car dependency: A worldwide analysis of car travel in global metropolitan areas.
\newblock {\em Cities}, 120:103467, 2022.

\bibitem{vanoutrive2015modal}
Thomas {Vanoutrive}.
\newblock The modal split of cities: A workplace-based mixed modelling perspective.
\newblock {\em Tijdschrift voor Economische en Sociale Geografie}, 106(5):503--520, 2015.

\bibitem{saeidizand2024patterns}
Pedram {Saeidizand} and Kobe {Boussauw}.
\newblock Patterns of car dependency of metropolitan areas worldwide: Learning from the outliers.
\newblock {\em International Journal of Sustainable Transportation}, 18(3):221--235, 2024.

\bibitem{schwanen2005affects}
Tim {Schwanen} and Patricia~L {Mokhtarian}.
\newblock What affects commute mode choice: neighborhood physical structure or preferences toward neighborhoods?
\newblock {\em Journal of Transport Geography}, 13(1):83--99, 2005.

\bibitem{maier2014dirichletreg}
Marco {Maier}.
\newblock Dirichletreg: {D}irichlet regression for compositional data in {R}.
\newblock Research Report 125, Vienna University of Economics and Business, 2014.

\bibitem{citeR}
{R Core Team}.
\newblock {\em R: A Language and Environment for Statistical Computing (version 4.2.3)}.
\newblock R Foundation for Statistical Computing, Vienna, Austria, 2014.

\bibitem{MacroTraffic25Cities}
Martín {Saavedra}, Alberto~P. {Muñuzuri}, Monica {Menendez}, and Jose {Balsa-Barreiro}.
\newblock Analysing macroscopic traffic rhythms and city size in affluent cities: insights from a global panel data of 25 cities.
\newblock {\em Philosophical Transactions A}, 382(2284):20240102, 2024.

\bibitem{de2019satisfying}
Jonas {De Vos}, Tim {Schwanen}, Veronique {Van Acker}, and Frank {Witlox}.
\newblock Do satisfying walking and cycling trips result in more future trips with active travel modes? an exploratory study.
\newblock {\em International Journal of Sustainable Transportation}, 13(3):180--196, 2019.

\bibitem{Tram19Cities}
Jan {Scheurer}.
\newblock How intermediate capacity modes provide accessibility and resilience in metropolitan transit networks: Insights from a global study of 19 cities.
\newblock {\em Journal of Public Transportation}, 19(4):107--125, 2016.

\bibitem{lindsey2023distributional}
Robin {Lindsey}, Ioannis {Tikoudis}, and Katherine {Hassett}.
\newblock Distributional effects of urban transport policies to discourage car use: A literature review.
\newblock {\em OECD Environment Working Papers}, 1(211), 2023.

\bibitem{GASCON2020362}
Mireia {Gascon}, Oriol {Marquet}, Esther {Gr{\`a}cia-Lavedan}, Albert {Ambr{\`o}s}, Thomas {G{\"o}tschi}, Audrey {de Nazelle}, Luc~Int {Panis}, Regine {Gerike}, Christian {Brand}, Evi {Dons}, et~al.
\newblock What explains public transport use? evidence from seven {E}uropean cities.
\newblock {\em Transport Policy}, 99:362--374, 2020.

\bibitem{ewing2017does}
Reid {Ewing} and Robert {Cervero}.
\newblock ``{D}oes compact development make people drive less?'' the answer is yes.
\newblock {\em Journal of the American Planning Association}, 83(1):19--25, 2017.

\bibitem{oviedo2022accessibility}
Daniel {Oviedo}, Clemence {Cavoli}, Caren {Levy}, Braima {Koroma}, Joseph {Macarthy}, Orlando {Sabogal}, Fatima {Arroyo}, and Peter {Jones}.
\newblock Accessibility and sustainable mobility transitions in {A}frica: Insights from {F}reetown.
\newblock {\em Journal of Transport Geography}, 105:103464, 2022.

\bibitem{hernandez2016transport}
Daniel {Oviedo Hernandez} and Julio~D {D{\'a}vila}.
\newblock Transport, urban development and the peripheral poor in {C}olombia—placing splintering urbanism in the context of transport networks.
\newblock {\em Journal of Transport Geography}, 51:180--192, 2016.

\bibitem{macedo2022differences}
Mariana {Macedo}, Laura {Lotero}, Alessio {Cardillo}, Ronaldo {Menezes}, and Hugo {Barbosa}.
\newblock Differences in the spatial landscape of urban mobility: gender and socioeconomic perspectives.
\newblock {\em Plos One}, 17(3):e0260874, 2022.

\bibitem{epomm}
{The European Platform on Mobility Management}.
\newblock {The European Platform on Mobility Management}, 2022.

\bibitem{UE_urbanaudit}
{European Union}.
\newblock {Urban Audit}, 2022.

\bibitem{ICLEI}
ICLEI.
\newblock {Local Governments for Sustainability - ICLEI}, 2022.

\bibitem{barmby1989modelling}
Tim {Barmby} and Jurgen {Doornik}.
\newblock Modelling trip frequency as a {P}oisson variable.
\newblock {\em Journal of Transport Economics and Policy}, pages 309--315, 1989.

\bibitem{mukherjee2022comprehensive}
Jaideep {Mukherjee} and B~Raghuram {Kadali}.
\newblock A comprehensive review of trip generation models based on land use characteristics.
\newblock {\em Transportation Research Part D: Transport and Environment}, 109:103340, 2022.

\bibitem{chang2014comparative}
Justin~S {Chang}, Dongjae {Jung}, Jaekyung {Kim}, and Taeseok {Kang}.
\newblock Comparative analysis of trip generation models: results using home-based work trips in the {S}eoul metropolitan area.
\newblock {\em Transportation Letters}, 6(2):78--88, 2014.

\bibitem{jang2005count}
Tae~Youn {Jang}.
\newblock Count data models for trip generation.
\newblock {\em Journal of Transportation Engineering}, 131(6):444--450, 2005.

\bibitem{kim2013comparison}
Nam~Seok {Kim} and Yusak~O {Susilo}.
\newblock Comparison of pedestrian trip generation models.
\newblock {\em Journal of Advanced Transportation}, 47(4):399--412, 2013.

\bibitem{UK_2020}
Department for~Transport UK.
\newblock {UK} {N}ational {T}ravel {S}urvey, 2020, 2020.

\bibitem{US_2017}
U.S. Department of~Transportation 2017 and Federal~Highway Administration.
\newblock {US} {N}ational {H}ousehold {T}ravel {S}urvey, 2017.

\end{thebibliography}

\end{document}